%% file: main.tex
\documentclass{article}
\usepackage{spconf,amsmath,graphicx}
\usepackage{amsmath,amsfonts}
\usepackage{algorithmic}
\usepackage{algorithm}
\usepackage{array}
\usepackage[caption=false,font=normalsize,labelfont=sf,textfont=sf]{subfig}
\usepackage{textcomp}
\usepackage{stfloats}
\usepackage{url}
\usepackage{verbatim}
\usepackage{graphicx}
\usepackage{cite}
\usepackage[center]{caption}
\usepackage{svg}

\usepackage{booktabs}
\usepackage{multirow}
\usepackage{multicol}

\usepackage{bbm}
\usepackage{breqn}
\usepackage{amsmath}

\title{On the Transferability of Large-Scale Self-Supervision to Few-Shot Audio Classification}
\name{Calum Heggan$^1$, Sam Budgett$^2$, Tim Hospedales $^1$, Mehrdad Yaghoobi$^1$\thanks{This work is supported by the Engineering and Physical
Sciences Research Council of the UK (EPSRC) Grant number
EP/S000631/1 and the UK MOD University Defence Research
Collaboration (UDRC) in Signal Processing, EPSRC iCASE account EP/V519674/1 and Thales UK Ltd. }}

\address{
  $^1$ University of Edinburgh, Scotland, 
  $^2$ Thales UK RTI}
\begin{document}
%
\maketitle
%

\begin{abstract}
    In recent years, self-supervised learning has excelled for its capacity to learn robust feature representations from unlabelled data. Networks pretrained through self-supervision serve as effective feature extractors for downstream tasks, including Few-Shot Learning. While the evaluation of unsupervised approaches for few-shot learning is well-established in imagery, it is notably absent in acoustics. This study addresses this gap by assessing large-scale self-supervised models' performance in few-shot audio classification. Additionally, we explore the relationship between a model's few-shot learning capability and other downstream task benchmarks. Our findings reveal state-of-the-art performance in some few-shot problems such as SpeechCommandsv2, as well as strong correlations between speech-based few-shot problems and various downstream audio tasks. 
    
\end{abstract}
\begin{keywords}
Self-Supervision, Few-Shot Learning
\end{keywords}
%
\section{Introduction}
Both few-shot learning and self-supervised learning have become increasingly popular in response to the lack of large labelled datasets in many domains and practical applications \cite{meta_survey,ssl_models_transfer}. Models pre-trained using self-supervision, the act of generating and solving self-generated tasks, have demonstrated strong success on few-shot learning tasks. Despite significant strides in other areas of benchmarking these approaches for audio problems questions persist regarding few-shot capabilities. Specifically, the effectiveness of self-supervision approaches for downstream few-shot adaptation is unclear due to the diverse range of methods and architectures employed in published models. This is further complicated by the substantial computational costs required to train large-scale models from scratch. The alignment of model rankings based on few-shot performance with those for other tasks is a critical consideration, influencing whether progressions from other audio-related tasks can be leveraged for few-shot learning. If misaligned, it may signal the need to include few-shot tasks in future holistic audio self-supervision benchmarks. 

\noindent To address these issues, we conduct an extensive evaluation of state-of-the-art (SOTA) pre-trained self-supervised models for downstream few-shot audio classification. Our study includes 13 pre-trained models evaluated across 10 diverse few-shot datasets, spanning environmental, animal, and speech sounds. Key contributions include, a) identifying the most effective approach for few-shot audio classification, b) understanding differences in algorithm ranking between few-shot and other benchmarks, and c) exploring relationships between few-shot and other tasks.

\section{Related Work}
\label{sec:related_work}

\subsection{Few-Shot Learning}
Few-shot learning aims to learn a task from limited labelled examples. Various fields address this challenge, with meta-learning being a prominent approach. Meta-learning involves generating and solving similarly structured few-shot tasks by leveraging a labelled training dataset. Two major groups emerge: I) Gradient-Based Meta-Learning (GBML), where models adapt rapidly to new tasks, and II) Metric-based approaches, which learn an embedding network\cite{meta_survey}. Classical pre-training with labelled base classes \cite{meta_survey} and, more recently, self-supervision, where models are trained on an unlabelled dataset using pre-text tasks, have also shown success \cite{ssl_models_transfer}. While algorithms for few-shot learning are prevalent in the image domain, their exploration in the audio domain is limited, with only a few being reproducible. We heavily rely on the MetaAudio benchmark \cite{metaaudio} and its extension from MT-SLVR \cite{mt-slvr} for our few-shot evaluation. This benchmark provides clear testing settings and diverse meta-learning paradigms for the evaluation tasks.

\subsection{Self-Supervision}
Self-supervised representation learning is a large and rich topic, both within acoustics and other modalities. As such we refer readers to more comprehensive surveys of available approaches and their unique attributes \cite{ssl_speech_survey}. Here, we instead focus on relevant families of algorithms. A popular family of approaches is prediction-based self-supervision, where a model is trained to predict the context of unseen sections of data \cite{apc,audio_albert,decoar_2,mockingjay,tera,wavlm}, or to contrast the unseen target frame with randomly sampled ones \cite{wav2vec, wav2vec2}. Multi-task approaches combining multiple objectives have also been explored \cite{paseplus,mt-slvr}. Discrete target methods, like clustering in HuBERT and DistilHuBERT \cite{hubert,distill_hubert}, also contribute to this rich landscape.

\subsection{Benchmarks \& Evaluations}
With increasing number of proposed approaches, the value of rigorous empirical evaluation has grown \cite{ssl_models_transfer}. Within few-shot classification, evaluation has largely been around imagery, however more recently has spanned into other domains. Of particular relevance to this work are the MetaAudio\cite{metaaudio} benchmark and its follow-up MT-SLVR \cite{mt-slvr}, which to date contain the most diverse set of downstream few-shot audio classification tasks. Within self-supervision, there has been a rapid growth of benchmarking efforts. Our work is most closely related to those focused on acoustics, the most ubiquitous of which is the Speech Processing Universal PERformance Benchmark (SUPERB) \cite{superb}. In total, SUPERB comprises 11 unique tasks but despite being diverse within the speech domain, lacks evaluation of few-shot capabilities, the ability to perform which is vital in many practical problems.

\input{tables/approaches_table_pre_trained}
\input{tables/fs_datasets}

\section{Self-Supervision For Few-Shot Learning}
\noindent
Self-supervised pre-training is a strong candidate for few-shot cases where a large quantity of unlabelled data is available. In such cases, we are able to learn general purpose representations which can then be used directly or updated with few samples. We study a pipeline in which trained models are frozen and used as feature extractors. Then for each task we train a lightweight linear classifier.  We evaluate the included approaches on few-shot audio classification tasks. Such few-shot tasks are individual learning problems, each containing a small training set (support set $\mathcal{S}$) and testing set (query set $\mathcal{Q}$). Few-shot tasks are commonly expressed as N-Way K-Shot tasks, where N is the number of classes being discriminated between, while K is the number of labelled examples per class available. Formally, N-Way K-Shot tasks take the form:
\begin{equation}
    \mathcal{S}=\left\{\left(x_{1}, y_{1}\right),\left(x_{2}, y_{2}\right), \ldots,\left(x_{\mathcal{M}}, y_{\mathcal{M}}\right)\right\}
    \label{eq:support_set}
\end{equation}
\begin{equation}
    \mathcal{Q}=\left\{\left(x_{1}, y_{1}\right),\left(x_{2}, y_{2}\right), \ldots,\left(x_{\mathcal{L}}, y_{\mathcal{L}}\right)\right\}
    \label{eq:query_set}
\end{equation}
where each $(x, y)$ pair consists of an input $\mathbf{x} \in \mathbb{R}^{D}$ and a class label $\mathbf{y} \in \{1, \ldots, N\}$, and where $\mathcal{M}$ and $\mathcal{L}$ are the total number of support and query examples respectively.

\input{results_tables/fs_results}
\section{Setup}
\label{sec:setup}

\subsection{Models \& Pre-Training}
We utilise models which have been submitted and evaluated on the SUPERB benchmark, and that are available through the s3prl toolkit \cite{tera,mockingjay}. To keep pipelines as fair as possible, we only evaluate models which have been pre-trained using LibriSpeech 960 (LS960). In total, we consider 13 models spanning a variety of training objectives, as detailed in Table \ref{table:approaches}. We refer the reader to the original works for implementation details. 

\subsection{Few-Shot Evaluation \& SUPERB}
We conduct evaluation of few-shot audio classification  across ten datasets, encompassing speech, environmental, and animal sounds (details in Table \ref{table:datasets}). We incorporate few-shot tasks beyond the speech pre-trained domain for two reasons: I) to assess the extent to which tasks based on animal sounds, somewhat similar to human speech, can be evaluated using speech features, and II) to gauge the out-of-domain transfer performance of speech models to other audio forms. For datasets designed for meta-learning, with class-wise splits, we exclusively use the provided test classes. This constraint ensures a fair comparison with previous works such as MetaAudio and MT-SLVR \cite{metaaudio, mt-slvr}. During evaluation, pre-trained models are frozen and employed as feature extractors, with new linear classifiers trained per task. All included models output 2D features, comprising both a traditional feature dimensions and a time dimension. Since this 2D representation isn't suitable for linear classifiers, we collapse one of the dimensions, with the method of doing so treated as a hyperparameter. We experimented with min, max and average pooling, and found that averaging the time dimension resulted in the highest validation performance. We also use the SUPERB benchmark, which comprises 11 tasks spanning 4 aspects of speech: content, speaker, semantics, and paralinguistics \cite{superb}.

\subsection{Correlation}
To analyse the relationship between downstream few-shot and upstream SUPERB benchmarks, we use  Pearson Rank correlation. To account for relative importance of absolute changes, we utilise the logit transformation of all metrics that have a range between 0 and 1 \cite{ssl_models_transfer}. This includes all accuracies and SUPERB measures except for model ranks score. Where possible, bootstrap resampled correlation errors are given.

\subsection{Limitations}
Due to availability of pre-trained models and compute resources, our work has limitations. Many SOTA audio methods often use unique architectures. This, together with the immense cost of training from scratch, means that our comparison cannot account for differences in backbone. We are also limited to linear readout evaluation due to fine-tuning cost.


\section{RESULTS}

\input{figures/corr_plots}

\input{figures/heatmap}

\subsection{Few-Shot Performance}
The few-shot results in Table \ref{table:fs_results} reveal a variety of noteworthy insights. Firstly, we note significantly improved performance in few-shot keyword spotting with SpeechCommandsV2 (SCv2) in almost all included models, with the highest performance being achieved by HuBERT Base. Intriguingly, pre-training with LS960, while effective for few-shot keywords, appears to yield suboptimal results for other included speech tasks compared to pre-training on environmental sounds \cite{mt-slvr}. For Kaggle18, NSynth and BirdClef, we achieve competitive or SOTA performance, an interesting result given that these sets are not based in speech. This result suggests that there is a possible overlap between required representation space for instrumentation/birdsong and speech. Although not competitive with SOTA approaches, performance on underwater mammal classification does overlap with many results from the joint training meta-learning condition in MetaAudio \cite{metaaudio}. Averaging over all sets, we observe that HuBERT base, followed closely by DistilHuBERT, performs best. Using average rank, we observe that APC performs consistently well, although suffering from poorer relative performance on a few key sets such as SCv2.

\subsection{Relationship to SUPERB}
We analyse the relationship between SUPERB and few-shot tasks by considering task-wise correlation Figures \ref{fig:corr_score} and \ref{fig:corr_heatmap}. We observe that speech sets, with the exclusion of VoxCeleb and Crema-D, exhibit significantly higher correlation across all tasks compared to environmental and animal sounds. Surprisingly, despite being rooted in speech, VoxCeleb and Crema-D display weak correlations with their traditional evaluation counterparts. This particularly interesting given that the same VoxCeleb dataset is used in SUPERBs' speaker identification task. In addition to its relative weakness, we also observe that VoxCeleb shows an unexpected negative relationship with considered tasks, e.g. as Automatic Speech Recognition performance gets worse, few-shot VoxCeleb performance improves. Interestingly, underwater mammals (Watkins) mirrors this behaviour. Environmental sets exhibit consistently low correlation with SUPERB tasks, while animal sounds achieve a moderately stronger relationship. For few-shot tasks that correlate well with any SUPERB task, correlation is strong across all tasks. The strongest single correlation is observed between Query-by-Example (QBE) and SCv2. Looking at the SUPERB benchmark more holistically, we can model the relationship between our few-shot tasks and SUPERB's  model score, Figure \ref{fig:corr_score}. Even in subdomain few-shot tasks such as few-shot speech, evidence suggests that performance is only aligned to SUPERB's scores in a few cases. For included speech problems, we note that performance variation over models is also fairly low compared to variation in SUPERB tasks, a possible extrapolation of which could suggests high specialisation in many of the included tasks gives only marginal gains on few-shot tasks. We also note that one reason for potential low correlation in many tasks is how features from these models are used downstream in SUPERB. Although the model itself is never fine-tuned, the linear model used on top of the frozen extractors typically has access to many more samples in SUPERB tasks than would be found in few-shot tasks. How different the features of a non data-constrained linear model for a given SUPERB task are from the base model features may have a significant impact. For robust future performance and low-data aligned SOTA models, firstly we propose that speech based few-shot tasks should be included in SUPERB style benchmarks for overall model scoring. In addition to adding speech based tasks, secondly we suggest that adding some additional related tasks such as animal sounds and instrumentation could be beneficial, due to their seemingly similar qualities.

\section{Conclusion}
In this study, we investigated the efficacy of large-scale self-supervised models in the realm of few-shot audio and speech classification, concurrently exploring their relationship with the widely-used self-supervised audio benchmark SUPERB. Our comprehensive evaluation of 13 models across 10 diverse few-shot audio datasets revealed notable insights, including the establishment of a new state-of-the-art for SCv2 and the limited impact of speech pre-training on few-shot speech tasks. Three out of five speech-based few-shot tasks demonstrated high correlations with SUPERB tasks, indicating potential domain connections. Animal sounds also exhibited some relation to SUPERB. Conversely, negative relationships were observed between many few-shot and SUPERB tasks, suggesting conflicts between few-shot performance and existing benchmarks. Our study implies that SUPERB benchmark performance improvements may not generalise to low-resource settings. Consequently, we propose the inclusion of few-shot tasks, especially speech-related ones, in future speech self-supervision benchmarks for a more comprehensive evaluation of model capabilities in diverse scenarios.

\newpage
\newpage
\bibliographystyle{IEEEbib}
\bibliography{refs}

\end{document}

%% file: tables/approaches_table_pre_trained.tex
\setlength{\tabcolsep}{4pt} 
\renewcommand{\arraystretch}{1.2} 
\begin{table}[htp]
\caption{Summary of pre-trained models. Abbreviations following \cite{superb}: (VQ) vector quantization, (F) future, (M) masked, (G) generation, (C) contrastive, (P)  token prediction/classification, (UM) utterance mixing and (GREP) gated relative position bias. Param count includes pre-training and inference.}
 \resizebox{1\columnwidth}{!}{%
\begin{tabular}{lcccc}
\toprule
Approach    &  $N^o$ Params (M) & Objective(s)    & Network &   Input           \\
\midrule
WavLM Base \cite{wavlm} &  94.38  & M-P + VQ + GREP + UM   & 7-Conv 12-Trans  &   Raw  \\
HuBERT Base \cite{hubert} & 94.68   &   M-P + VQ  & 7-Conv 12-Trans  &   Raw \\
wav2vec 2.0 Base \cite{wav2vec2} & 95.04   & M-C + VQ   & 7-Conv 12-Trans &   Raw  \\
DistilHuBERT \cite{distill_hubert} &  21.32  &  Layer Distillation  & 7-Conv 2-Trans  &  Raw  \\
DeCoAR 2.0 \cite{decoar_2} &  89.84  &  M-G + VQ  & 12-Trans &   Raw  \\
wav2vec \cite{wav2vec} &  32.54  & F-C   & 19-Conv  &  Raw  \\
vq-wav2vec \cite{vq_wav2vec} & 34.15   &   F-C + VQ & 20-Conv  &   Raw \\
APC \cite{apc} & 4.11   & F-G   & 3-GRU  &  Raw  \\
VQ-APC \cite{vq_apc} & 4.63   & F-G + VQ   & 3-GRU  &  Raw  \\
NPC \cite{npc} & 19.38   &  M-G + VQ  & 4-Conv, 4-Masked Conv  &   Raw \\
TERA \cite{tera} &  21.33  &  Time/Freq M-G  & 3-Trans  &   Spec \\
PASE+ \cite{paseplus} & 7.83   & Multi-Task   &  SincNet, 7-Conv, 1-QRNN &   Raw \\
MockingJay \cite{mockingjay} &  21.33  &  Time M-G  & 12-Trans &   Spec  \\


\bottomrule
\end{tabular}}
\label{table:approaches}
\end{table}

%% file: tables/fs_datasets.tex
\setlength{\tabcolsep}{6pt} 
\renewcommand{\arraystretch}{1} 
\begin{table}[htp]
  \caption{Summary of few-shot  datasets: Speech (top), environmental (middle), and animal (bottom) sounds.}
  \centering
   \resizebox{\columnwidth}{!}{%
  \begin{tabular}{ccccc}
    \toprule
    Name & Setting & $N^o$ Classes & $N^o$ Samples & Length (s) \\
    \midrule

    VoxCeleb1   & Speaker  & 1,251 & 153,516  & 3-180\\
    SpeechCommandsV2    & Keyword & 35 & 105,829    & 1\\
    Crema-D  & Emotion Recognition & 6 & 7,442 & 1s - 5\\
    Speech Accent Archive  & Accent& 122 & 2,060  & 17s - 110\\
    Common Voice v12 Delta  & Language & 88 & 256,243  & 530\\
    \midrule 
    
    ESC-50        & Environmental         & 50   & 2,000       & 5\\
    NSynth      & Instrumentation       & 1,006 & 305,978     & 4\\
    FDSKaggle18   & Mixed                 & 41   & 11,073  & 0.3-30\\
    \midrule
    Watkins Marine Mammal       & Marine Mammals         & 32   & 1,698       & 0.1-150\\
    BirdCLEF 2020 (Pruned)  & Bird Song    & 715  & 63,364  & 3-180 \\

    \bottomrule
  \end{tabular}}
  \label{table:datasets}
\end{table}


    


%% file: results_tables/fs_results.tex
\setlength{\tabcolsep}{6pt} 
\renewcommand{\arraystretch}{0.95} 
\begin{table*}[htp]
    \centering
    \caption{Average percentage accuracy for  few-shot audio classification. Each result is the mean and 95\% confidence interval of 10,000 random 5-way 1-shot tasks. We also include current SOTA results. Results style: \underline{\textbf{Best}}, \textbf{Second Best}.
}\label{table:fs_results}
    \resizebox{2\columnwidth}{!}{%
        \begin{tabular}{c|ccccc|ccc|cc|cc}
            \toprule
           Method & \multicolumn{5}{c}{Speech} & \multicolumn{3}{c}{Enviromental} & \multicolumn{2}{c}{Animal} & \\
            \midrule
             & SCv2 & SAA & CommonVoice & VoxCeleb & Crema-D & NSynth & ESC-50 & Kaggle18 & Watkins & BirdClef & Avg & Avg Rank\\
            \midrule
            WavLM Base & 52.27$^{\pm0.43}$ & \textbf{26.92$^{\pm0.33}$} &\textbf{31.72$^{\pm0.38}$} & 27.68$^{\pm0.36}$ & 27.86$^{\pm0.38}$ & 56.69$^{\pm0.42}$ & 48.29$^{\pm0.40}$ & 35.85$^{\pm0.40}$ & 43.29$^{\pm0.42}$ & 28.26$^{\pm0.37}$ & 37.88$^{\pm0.07}$ &7.1\\
            
            HuBERT Base & \underline{\textbf{57.10$^{\pm0.43}$}} & 26.30$^{\pm0.34}$ & 31.34$^{\pm0.38}$ & 28.36$^{\pm0.37}$ & 28.17$^{\pm0.38}$ & 61.45$^{\pm0.41}$ & \textbf{55.41$^{\pm0.41}$} & 37.71$^{\pm0.41}$ & 44.25$^{\pm0.43}$ & 30.30$^{\pm0.38}$ & \textbf{40.04$^{\pm0.07}$} & 4.7\\
            
            wav2vec 2.0 Base & 34.41$^{\pm0.39}$ & 25.67$^{\pm0.34}$ & 30.02$^{\pm0.37}$ & 27.30$^{\pm0.36}$ & \underline{\textbf{29.91$^{\pm0.37}$}} & 50.09$^{\pm0.42}$ & 46.90$^{\pm0.42}$ & 33.20$^{\pm0.39}$ & 41.34$^{\pm0.42}$ & 28.04$^{\pm0.37}$ & 34.69$^{\pm0.06}$& 9.5\\
            
            DistilHuBERT & \textbf{55.20$^{\pm0.43}$} & 25.98$^{\pm0.34}$ & 31.63$^{\pm0.38}$ & 28.27$^{\pm0.37}$ & 28.80$^{\pm0.38}$ & 60.17$^{\pm0.41}$ & 55.16$^{\pm0.41}$ & 36.60$^{\pm0.41}$ & 45.47$^{\pm0.42}$ & 29.70$^{\pm0.37}$ & \textbf{39.70$^{\pm0.07}$} &4.9\\
            
            DeCoAR 2.0 & 37.05$^{\pm0.40}$ & 24.19$^{\pm0.34}$ & 30.32$^{\pm0.38}$ & 30.62$^{\pm0.38}$ & 26.72$^{\pm0.37}$ & 66.95$^{\pm0.39}$ & 49.14$^{\pm0.41}$ & 34.24$^{\pm0.38}$ & 43.23$^{\pm0.42}$ & \underline{\textbf{30.87$^{\pm0.37}$}} & 37.33$^{\pm0.06}$ &7.4\\
            
            wav2vec & 41.08$^{\pm0.41}$ & 22.15$^{\pm0.33}$ & 30.88$^{\pm0.38}$ & 28.25$^{\pm0.37}$ & 27.83$^{\pm0.38}$ & 49.83$^{\pm0.42}$ & 51.21$^{\pm0.41}$ & 34.58$^{\pm0.39}$ & 42.32$^{\pm0.41}$ & 28.00$^{\pm0.37}$ & 35.61$^{\pm0.05}$& 9.4\\
            
            vq-wav2vec & 41.06$^{\pm0.41}$ & 22.04$^{\pm0.31}$ & 27.88$^{\pm0.37}$ & 26.60$^{\pm0.36}$ & 28.86$^{\pm0.37}$ & 50.40$^{\pm0.41}$ & 48.34$^{\pm0.39}$ & 32.36$^{\pm0.38}$ & 37.76$^{\pm0.42}$ & 27.78$^{\pm0.37}$ & 34.31$^{\pm0.07}$ &11.2\\
            
            APC & 42.01$^{\pm0.41}$ & 22.42$^{\pm0.34}$ & 31.01$^{\pm0.38}$ & \textbf{32.47$^{\pm0.39}$} & 29.45$^{\pm0.38}$ & 64.38$^{\pm0.40}$ & 52.77$^{\pm0.42}$ & 36.12$^{\pm0.40}$ & 46.28$^{\pm0.43}$ & \textbf{30.50$^{\pm0.37}$} & 38.74$^{\pm0.06}$&  \textbf{4.5}\\
            
            VQ-APC & 38.95$^{\pm0.40}$ & 24.72$^{\pm0.34}$ & 29.46$^{\pm0.37}$ & 29.83$^{\pm0.38}$ & 28.56$^{\pm0.37}$ & 63.37$^{\pm0.40}$ & 49.89$^{\pm0.41}$ & 34.50$^{\pm0.38}$ & 43.23$^{\pm0.41}$ & 27.05$^{\pm0.35}$ & 36.96$^{\pm0.07}$ &8.4\\
            
            NPC & 31.18$^{\pm0.37}$ & 21.65$^{\pm0.31}$ & 27.55$^{\pm0.35}$ & 28.40$^{\pm0.35}$ & 27.49$^{\pm0.35}$ & 59.44$^{\pm0.40}$ & 46.26$^{\pm0.40}$ & 33.05$^{\pm0.37}$ & 43.12$^{\pm0.41}$ & 27.90$^{\pm0.35}$ & 34.60$^{\pm0.05}$&11.6 \\
            
            TERA & 32.56$^{\pm0.43}$ & 24.30$^{\pm0.33}$ & 30.21$^{\pm0.38}$ & 30.47$^{\pm0.36}$ & 29.24$^{\pm0.38}$ &69.45$^{\pm0.42}$ & 52.47$^{\pm0.40}$ & 35.11$^{\pm0.40}$ & \textbf{50.68$^{\pm0.42}$} & 29.85$^{\pm0.37}$ & 38.43$^{\pm0.05}$& 5.6\\
            
            PASE+ & 30.00$^{\pm0.37}$ & 24.70$^{\pm0.34}$ & 30.76$^{\pm0.36}$ & 26.86$^{\pm0.34}$ & 26.50$^{\pm0.36}$ & 37.36$^{\pm0.39}$ & 52.20$^{\pm0.40}$ & \underline{\textbf{39.20$^{\pm0.40}$}} & 45.30$^{\pm0.42}$ & 28.66$^{\pm0.34}$ & 34.15$^{\pm0.07}$ &8.7\\
            
            Mockingjay & 29.79$^{\pm0.38}$ & 23.58$^{\pm0.33}$ & 29.65$^{\pm0.37}$ & 28.87$^{\pm0.37}$ & 28.61$^{\pm0.38}$ & \textbf{70.71$^{\pm0.38}$} & 46.28$^{\pm0.42}$ & 33.52$^{\pm0.39}$ & 47.23$^{\pm0.43}$ & 27.61$^{\pm0.36}$ & 36.58$^{\pm0.05}$& 8.9\\

            \midrule
            MT-SLVR (SOTA) \cite{mt-slvr} & 23.65$^{\pm0.34}$ & \underline{\textbf{28.92$^{\pm0.37}$}} & \underline{\textbf{35.22$^{\pm0.40}$}} & \underline{\textbf{33.58$^{\pm0.39}$}} & \textbf{29.61$^{\pm0.38}$} & \underline{\textbf{71.81$^{\pm0.39}$}} & \underline{\textbf{69.53$^{\pm0.39}$}} & \textbf{38.36$^{\pm0.40}$} & \underline{\textbf{59.49$^{\pm0.42}$}} & 29.49$^{\pm0.38}$ & \underline{\textbf{41.97$^{\pm0.02}$}} & \underline{\textbf{3.0}} \\
            \bottomrule
        \end{tabular}
        }
        
\end{table*}

%% file: figures/corr_plots.tex
\begin{figure*}[htp]
    \centering
    \includegraphics[width=1\linewidth]{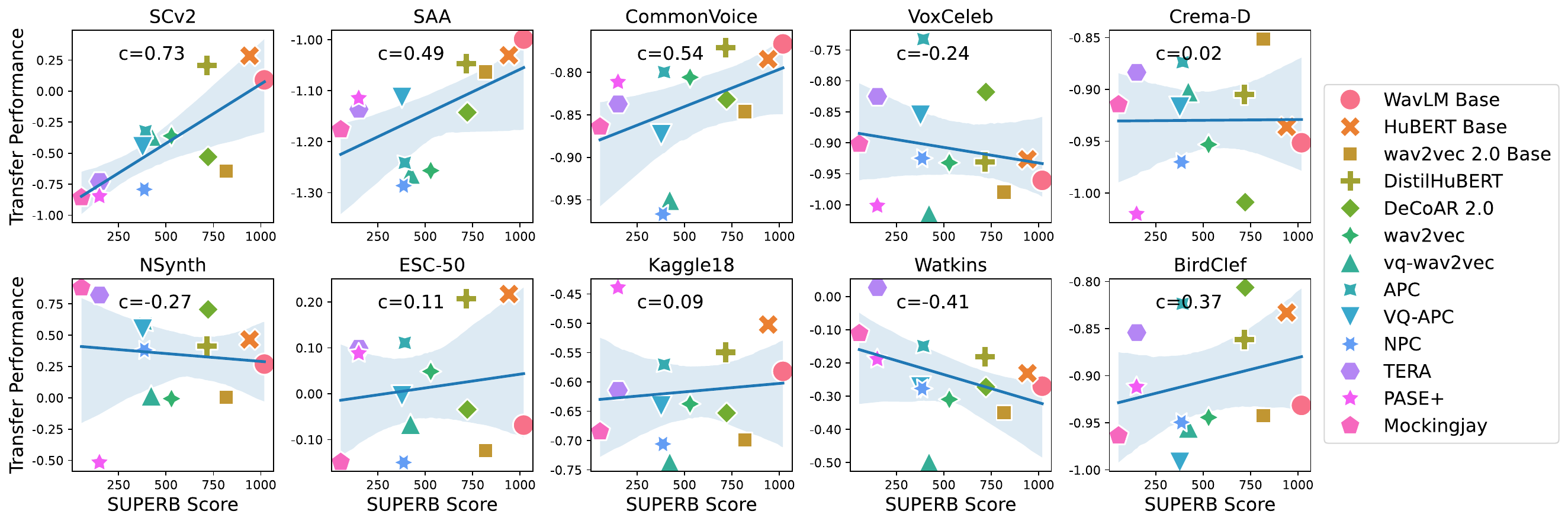}
    \caption{SUPERB model score vs average few-shot transfer performance for all considered datasets. (TOP) row contains speech datasets, (BOTTOM) row contains environmental/animal sets. Regression gradients and shaded regions describe correlation strength and 95\% confidence intervals respectively. Spearman Rank correlation coefficients (c) are shown top left of each plot.}
    \label{fig:corr_score}
\end{figure*}

%% file: figures/heatmap.tex

\begin{figure}[ht]
    \centering
    \resizebox{0.8\columnwidth}{!}{\includegraphics{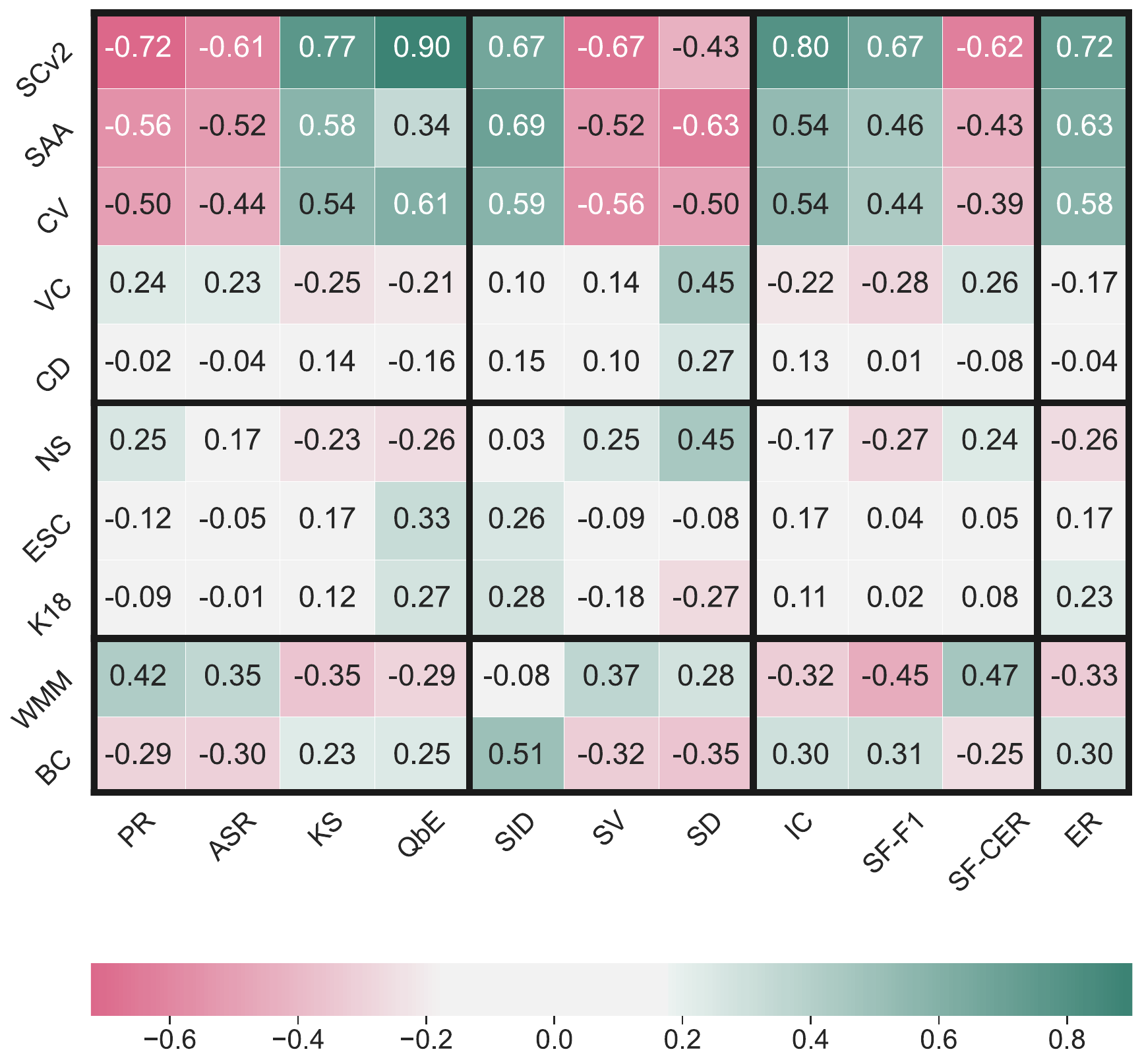}}
    \caption{Spearman rank correlations between Few-Shot (rows) and SUPERB (cols) tasks. Few-shot tasks are split into speech (top), environment (mid) and animal (bottom) sounds. SUPERB is split into context, speaker, semantics and paralinguistics  (left to right). 
    }
    \label{fig:corr_heatmap}
\end{figure}